\documentstyle[12pt,colortab,epsfig]{article}
\begin{document}
\vspace*{-1in}
\renewcommand{\thefootnote}{\fnsymbol{footnote}}
\begin{flushright}
 MRI-PHY/P990512 \\
\end{flushright}
\vskip 5pt
\begin{center}
{\Large{\bf $\nu_\tau$ Appearance Searches using Neutrino Beams \\
from Muon Storage Rings }}
\vskip 25pt
{\sf Sukanta Dutta $^{a,b\!\!}$
\footnote{E-mail address: sukanta@mri.ernet.in}},
{\sf Raj Gandhi $^a$ }
\footnote{E-mail address: raj@mri.ernet.in  } 
and 
{\sf Biswarup Mukhopadhyaya $^a$}
\footnote{E-mail address: biswarup@mri.ernet.in }   
\vskip 10pt
$^a${\it
Mehta Research Institute, Jhusi, Allahabad, India
}\\
$^b${\it
S.G.T.B. Khalsa College, Univ. of Delhi, Delhi, India
} \\

\vskip 20pt
\end{center}
\begin{abstract}
We study the possibilities offered by muon storage rings 
 for $\nu_\tau$ appearance
experiments in order to determine masses and mixing angles for $\nu_\mu
\rightarrow \nu_\tau$ and $\bar\nu_e
\rightarrow \bar\nu_\tau$ oscillations. Tau event rates for such
experiments are first  discussed with a view to examining  their 
variation prior to the inclusion of experimental cuts, in order to 
better understand how baselines, forward peaking of decay 
neutrinos with increasing energies,
and average fluxes intercepted by detectors of
 various sizes can affect their optimization. 
Subsequently, event rates implementing cuts for hadronic and 
wrong sign leptonic modes are computed and used to plot $90\%$ CL contours for 
the parameter regions that can be explored in such experiments, and the 
expected scaling of the contours with energy and baseline is discussed.
 The results show that even for modest muon beam energies,
 convincing coverage and verification of the 
Super Kamiokande  parameters is possible. In addition, very significant
enlargement of present day bounds on the mixing parameters for neutrino
oscillations of all types  is   guaranteed by these types of searches.
\end{abstract}
\begin{quotation}
{\small  PACS number(s): }
\end{quotation}

\vskip 20pt  

\setcounter{footnote}{0}
\renewcommand{\thefootnote}{\arabic{footnote}}
\vfill
\eject
\setcounter{page}{1}
\pagestyle{plain}
\section{Introduction}  The recent results of the Super Kamiokande
(Super K) water Cerenkov-detector experiment \cite{fuk} 
provide firm indications
of an anomaly in the flavor ratios and zenith-angle dependence of the
atmospheric neutrino flux. Although the existence of such an anomaly 
had already been signalled
by earlier data from the Kamiokande \cite{kam} and IMB \cite{imb}
experiments and supported by subsequent Soudan II results \cite{sou}, the
impressive statistical significance of the Super K data has
appreciably buttressed its interpretation in terms
neutrino mass and oscillations. This is especially true of the
observed zenith angle dependence of the observations, which does 
not naturally
seem to lend itself to any alternative explanation. When combined
with results from the CHOOZ reactor experiment \cite{chooz}, analyses
\cite{fog,nar} of the data tilt the balance towards an interpretation 
in terms of
$\nu_\mu \longrightarrow \nu_\tau $ oscillations  versus other
explanations. Evidence that this channel is favoured also comes
from neutral current event count ratios  involving the production of
neutral pions measured at Super K \cite{kaj}.

\par In addition to being the first firm signal for physics beyond 
the Standard
Model, a determination, even if approximate,  of neutrino masses
and mixing angles would be  a crucial pointer towards the nature of 
such physics, providing an unprecedented glimpse into what lies 
beyond present 
knowledge of particle interactions. Thus, the importance of independently
verifying the presence of $\nu_\mu \longrightarrow \nu_\tau $
oscillations can scarcely be overestimated. The firmest confirmation
of this hypothesis would be  via  the detection of $\tau$ leptons
produced by charged current interactions of $\nu_\tau $'s resulting
from oscillations of $\nu_\mu$'s. In this paper, we study this possibility
in the context of neutrinos obtained from muon storage rings at future 
muon colliders.

\par At present, high energy ( $\geq $ GeV) neutrino beams for
oscillation studies are obtained by allowing charged pions and kaons
produced in  fixed target accelerator experiments to decay in
flight. Recently, however, a new type of neutrino beam, much more
intense than those presently available, has been proposed 
 and
discussed for neutrino oscillation studies and other neutrino related
experiments \cite{barger,ellis,quig,geer,king,geer2,bueno1,bueno2,deruj}. 
These beams originate from a high
intensity muon source, currently under active design and study as part
of an effort to develop a high luminosity muon collider \cite{anken}. In addition
to the extremely intense and collimated primary neutrino fluxes which
will be available from such a source, the beam compositions will be
much more precisely known than in those available from pion and kaon
decay. A muon storage ring with the straight section pointing towards
a neutrino detector situated at a specific baseline length, as
described in detail in \cite{geer}, would lead to a neutrino beam with
precisely equal numbers of $\nu_\mu$ and $\bar\nu_e$, or alternatively
$\bar\nu_\mu$ and $\nu_e$,  depending on the sign of parent
muons. This is in contrast to the presently available high energy
neutrino beams from accelerators, which contain mostly muon neutrinos,   
 but with
small contaminations of electron and tau neutrino species. For $\nu_\tau$
appearance searches, the vastly superior luminosities, absence of
contamination and the possibility of higher energies of muon
collider neutrino beams make them an attractive proposal which
merits further study.
\par Accordingly,  we focus here  on the physics 
 of tau appearance experiments using neutrinos from muon 
storage rings,  depicting the two flavour
oscillation parameter ranges ($ \Delta m^2$ and
$\sin^22\theta $) which can be consequently  probed  in a
search for $\nu_\mu \longrightarrow \nu_\tau $ and  $\nu_e
\longrightarrow \nu_\tau $  oscillations.
\footnote{A full study would involve three flavour oscillations, but
our purpose here is to explore the possibilities rather than details,
given the fact that muon colliders and neutrino experiments fueled
by muons from storage rings  are still in the design and
development  stage; see for instance \cite{anken}.}
\par Adopting the  sample  design configuration  for muon production, 
capture,
cooling, acceleration and storage prior to decay described in \cite{geer},
 the number of available muons of either sign is $ \approx 8 \times 10^{20}$ per year.
Of these, one fourth decay in a straight section directed towards the 
neutrino detector, yielding $2 \times 10^{20}$ neutrinos
and an identical number of anti-neutrinos $( \nu_{\mu}$ and $ \bar{ \nu_e}$, if, for
example, the beam is comprised of $\mu^-).$
We use these numbers in all of the following calculations, and refer the reader
to \cite{geer} for design details leading to the production of the neutrino beams.

\par In section 2 we discuss the broader physics characteristics and
dependences of $\tau$ production rates at such oscillation experiments.
 In section 3 we
discuss  the realistic detection of $\tau$ events above
backgrounds. The channels we study are the detection of (i)
$\nu_\mu \longrightarrow \nu_\tau $ oscillations   via charge current
production and subsequent decay into hadrons, and (ii) $\bar\nu_e
\longrightarrow \bar\nu_\tau $  oscillations  via  the appearance
of wrong sign muons from $\tau$ decay to leptonic modes. The specific
conventional kT type detector,  discussed recently in  \cite{geer} is considered, and
we describe our choices for the kinematic cuts and/or overall
detection efficiencies for it. 

 In section  4 we use the results of the
consequent event rate calculations to present 90$\%$ CL contours for 
$ \Delta m^2 $ and $ \sin^22\theta $
 for a variety of
muon beam energies and  baseline lengths  in order to
illustrate the extraordinary possibilities offered by muon colliders
for studying neutrino oscillations.

\section{$\tau^-$ appearance event rates using muon collider beams:
general characteristics}

For the general discussion that this section focuses  on, we compute and use  
the actual
$\nu_\tau\longrightarrow \tau$ charged current (CC) production rates without including
experimental cuts to eliminate backgrounds. These are detailed and
incorporated later,  in sections 3 and  4, prior to 
obtaining contour plots for
$\Delta m^2$ and $\sin^22\theta$.
\par The event rate $N_\tau$ (events/kT/year) for $\tau$ lepton
production from $\nu_\mu$'s subsequent to oscillation is given by
\begin{equation}
N_\tau= 6.023\times 10^{32}\,\int {\sigma^{CC}_{\nu_\tau}\, {\cal
P}_{\nu_{\mu,e}\rightarrow\nu_\tau} \, {d\left\langle\Phi_{\nu_\mu}
\right\rangle\over dE_\nu}\, dE_\nu} \label{eq1}
\end{equation}
 where $\sigma_{\nu_\tau}^{CC}$ is the total charged current cross section
 obtained by integrating equation (\ref{eq3}) below. The oscillation 
probability  between flavours is 
\begin{equation}
{\cal P}_{\nu_{\mu,e}\rightarrow \nu_\tau} = \sin^22\theta\,\, \sin^2\left[ 
1.27\, \Delta m^2\, {L\over E_\nu}\right] \label{eq2}
\end{equation}
with $\Delta m^2= m^2_{\nu_\tau}\,- m^2_{\nu_{\mu,e}}$ in 
eV$^2$, $ L= $ baseline
length in km,
 $E_\nu$ being the neutrino energy in GeV, and $\theta$  the 
mixing angle between
flavours. $\left\langle\Phi_{\nu_\mu}\right\rangle $ is the number of
neutrinos in the cone intercepted by the detector averaged over its
area. The numerical factor is the number of scatterers ( iso -scalar
nucleons) per kT of the detector material.

\par  We first discuss the cross-section, performing  our calculation within 
the  renormalization group improved 
parton model, and focus on the inclusive process
$\nu_\tau N \longrightarrow \tau^- \, + $ anything, where $N$ is an 
isoscalar nucleon.
 On retaining effects of the $\tau$-mass, \footnote{ Or $\mu$-mass, in the case of 
$\sigma_{\nu_\mu}^{CC}$, which is also calculated below. Clearly, however, the 
terms proportional to the square of the lepton mass in Eqs. (3) and (4)
are important at the energies of interest for only the $\tau$-lepton.}  
the differential cross section can be written in terms of
the Bjorken scaling variables $x=Q^2/2 M\nu$ and $ y= \nu /E_\nu$ as
\begin{eqnarray}
{d^2\sigma^{\nu,\bar\nu}\over dx\,dy}&=& {G_F^2\, M\, E_\nu\,\over \pi}\;
\Bigg[ \left\{ xy + {m_\tau^2\over 2\, M\, E_\nu} \right\} \,F_1
 +\left\{ \left(1-y\right)-\left( {M\over 2\,E_\nu}xy\,+{ m_\tau^2\over
4\, E_\nu^2}\right)\right\} F_2\nonumber\\
&& \mp \left\{ xy\left(1- {1\over 2}
y\right)- {m_\tau^2\over 4\,M\, E_\nu}y\right\} F_3 \, +
 {m^2_\tau\over M^2}\left\{ \left( {M\over 2\, E_\nu}xy + {m_\tau^2\over
4\, E_\nu^2}\right)\, F_4 \, - {M\over 2\, E_\nu}\,F_5\right\}\Bigg]\nonumber\\
&& \label{eq3}
\end{eqnarray}
Here $- Q^2$ is the invariant momentum transfer between the incident
neutrino and outgoing tau, $\nu =E_\nu - E_\tau$ is the energy loss in
the lab (target) frame, $M$ and $M_W$ are the nucleon and intermediate
boson masses respectively, and $G_F=1.16632\times 10^{-5}$ 
GeV$^{-2}$ is the Fermi
constant. 
The limits on $x$ and $y$ are 
$${m_\tau^2\over 2\, M\left(E_\nu-M\right)}\leq x\leq 1\,, \hskip 2 cm
A-B\leq  y\leq A+B $$
where 
\begin{eqnarray}
A&=&{1\over2}\left( 1-{m_\tau^2\over 2\,M\,E_\nu x}-{m_\tau^2\over
2\,E_\nu^2}\right)/\left( 1+x{M\over 2 E_\nu}\right),\nonumber \\
B&=&{1\over2}\left\vert\left[\left( 1-{m_\tau^2\over 2\,
M\,E_\nu x}\right)^2  -
{m_\tau^2 \over 2\,E_\nu^2} \right]\right\vert^{1\over2}/\left( 
1+x{M\over 2
E_\nu}\right).\label{eq4}
\end{eqnarray}
The  $F_{i}$'s are given as
\begin{eqnarray}
F_2&=& x\,\left( {\it u} +{\it d}+\bar{\it u} + \bar{\it d} +2\, {\it s}
+ 2\, \bar{\it b}\right),\nonumber\\
F_1&=&{F_2\over 2\, x},\nonumber\\
F_3&=&\left({\it u} +{\it d}-\bar{\it u} - \bar{\it d} +2\, {\it s}
- 2\, \bar{\it b}\right),\nonumber\\
F_4&=& \left( -{M\,\nu\over Q^2}\right)^2\, F_2 + \left( -{M\,\nu\over Q^2}
\right)\, F_1,\nonumber \\
F_5&=& {F_2\over x}\nonumber
\end{eqnarray}

A  straightforward application of the Callan-Gross equations shows that $F_4$
 vanishes in this case. In the above, {\it u,d,c,s} and {\it b}  denote 
the distributions for the various
quark flavours in a proton. For our calculations we use CTEQ4 parton
distributions \cite{cteq}.

\par
\begin{figure}[htb]
\vspace*{-1.9 in}
\centerline{\epsfig{file=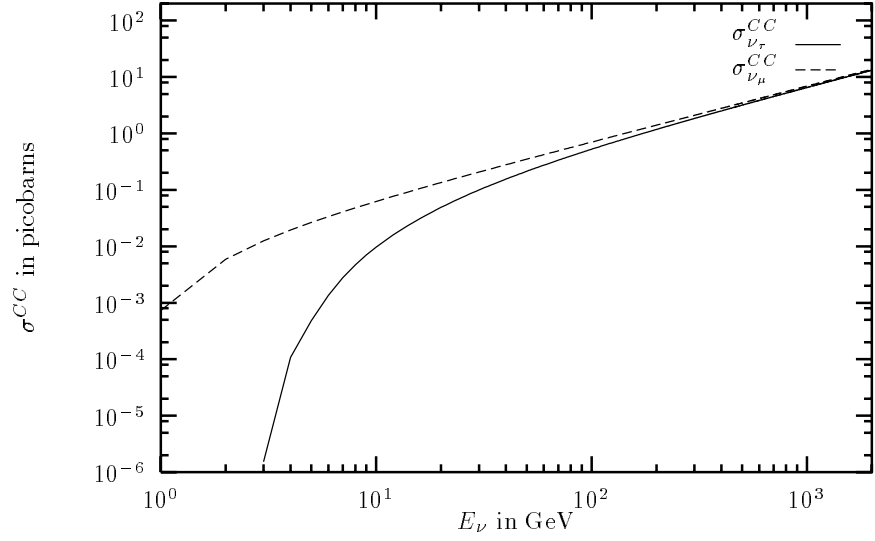,width=20cm}}
\vspace*{-6.5 in}
\caption{ \em The CC $\nu_\tau$
  - nucleon and $\nu_\mu$-nucleon  scattering cross sections.  } \label{fig1}
\end{figure}

 Figure \ref{fig1} shows the total CC 
cross sections for $\nu_\mu$ and
$\nu_\tau$, obtained using the above expressions. For convenience, we 
give analytic fits for both cross-sections in Table \ref{table1}, 
over the entire range of energy which is  
of interest here.

\par
\begin{figure}[htb]
\vspace*{-1.9 in}
\centerline{\epsfig{file=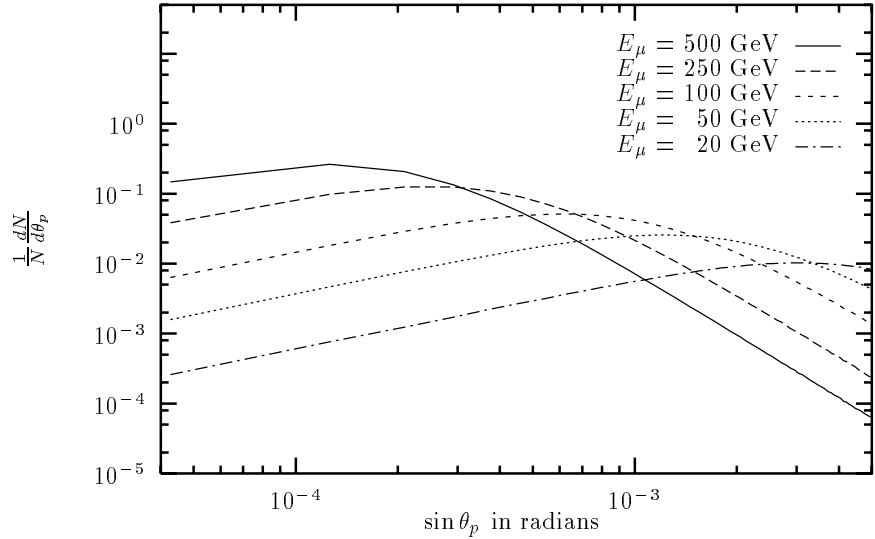,width=20cm}}
\vspace*{-6.5 in}
\caption{\em The angular distribution with  polar angle $\theta_p$
(angle  between the  initial muon beam and decaying muon neutrino) 
for various muon beam energies.   } \label{fig2} 
\end{figure}

 For the calculations of event rates,  we assume that area
dimensions of the detector and the baseline length $L$ define a \,\lq detection cone\rq\,\, of half angle
$\theta_d $ with the direction of the muon beam, with detector radius
$R_d\equiv L\theta_d$. Thus, for long baselines the choice of
$\theta_d$ would expectedly be  smaller than those for shorter 
baselines in order to
accommodate a realistic detector size. The angular distribution of
$\nu_\mu$ within a chosen detection cone, of course, follows from the
decay kinematics of muon, and is obtained by boosting the familiar
distribution of a muon decaying at rest to the requisite beam
energy. Figure \ref{fig2} shows, for various beam energies, the angular
distribution ${1\over N_{\nu_\mu}} \, {d N_{\nu_\mu}(\theta_p) 
\over d\theta_p}$
in the polar angle $\theta_p$,
 where $N_{\nu_\mu}(\theta_p) $ is the number of muon neutrinos  ( prior
to any oscillation ) contained within a  cone of half-angle  $\theta_p$,
demonstrating the expected forward peaking of muon neutrinos with
increasing parent particle energy.  This distribution peaks around $1.2 \times 10^{-4}$ radians for $E_{\mu} = 500$ GeV, around $3 \times 10^{-4}$ radians for 
$E_{\mu}= 250$ GeV, and even higher for $E_{\mu}=$ 50 and 20 GeV.
We remark here that for the wrong-sign muon detection mode discussed 
and used below, the parent $\bar{\nu_e}$ angular distribution differs 
from the $\nu_\mu$
distribution shown here, as dictated by decay kinematics.

\par
\begin{figure}[htb]
\vspace*{-1.9 in}
\centerline{\epsfig{file=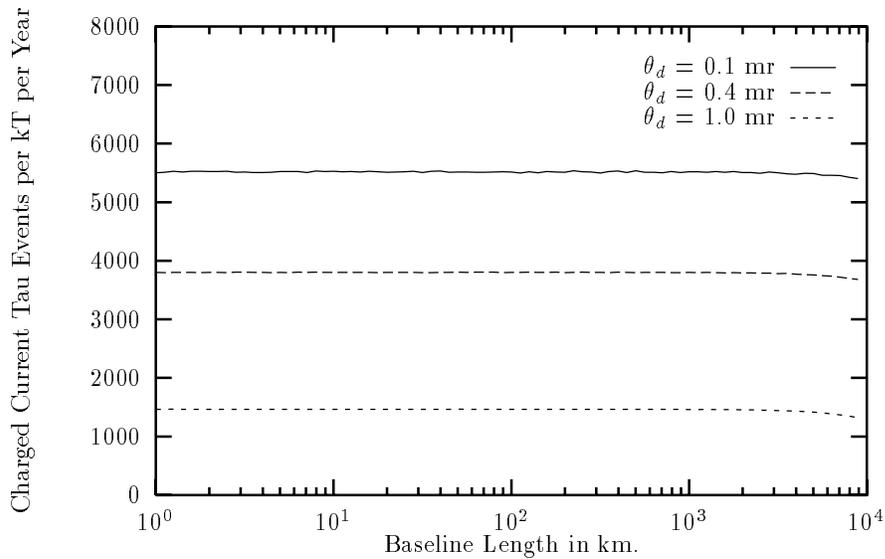,width=20cm}}
\vspace*{-6.6 in}
\caption{\em The insensitivity of the tau event rate to variations in baseline
for a fixed detection cone, as discussed in the text. Values of the oscillation parameters are chosen in accordance with the best fits   from
Super K results.} \label{fig3} 
\end{figure}

 Next, we note that the oscillation probability in equation
(\ref{eq2}) reduces to 
\begin{equation}
{\cal P}_{\nu_\mu\longrightarrow\nu_\tau} \simeq 1.27^2\,\sin^22\theta \,
\Delta m^4\, {L^2\over E_\nu^2}\label{eq5}
\end{equation}
for $\Delta m^2 L/E_\nu \leq 0.1$,
and that this condition is satisfied for a significant range of $\Delta
m^2$ values, even for large baseline lengths, given the high energies
contemplated for muon colliders. It is, for instance, valid for the
favoured value of Super K results, $\Delta m^2\approx 10^{-3}$ eV$^2$, 
$\sin^22\theta \approx 1$, 
even for a relatively low 20 GeV muon beam and a 732 km
baseline. Since the average flux intercepted by a detector for some
fixed $\theta_d$ falls as $1/L^2$, from equation (\ref{eq1}) we see that
$N_\tau$ will be independent of baseline length as long as 
equation (\ref{eq5}) is satisfied. This independence is illustrated in 
figure \ref{fig3}, where
the tau events /kT/yr are plotted versus baseline lengths for three
different values of $\theta_d$, for a 250 GeV muon beam energy and
$\Delta m^2 = 10^{-3}$ eV$^2$.

\par Subsequent to oscillation, and when  the detector area is taken 
into account, the enhanced collimation of the neutrinos with increasing 
beam energy   manifests itself
  in an interesting manner.  In general, for a significant part of the range 
of the energies of interest at muon colliders, 
$\sigma_{\nu_{\tau}}^{CC}$ rises linearly
with neutrino energy. For our choice of oscillation parameters, 
the probability
${\cal P}_{\nu_\mu\longrightarrow \nu_\tau}$ varies with energy as 
$E_\nu^{-2}$ in equation (\ref{eq5}) , while the forward peaking of the neutrino 
beam with energy enhances the flux term as $E_\nu^{2}$, leading to an 
overall linear increase in the event rate with energy for a detector of 
fixed mass whose area matches that of the kinematic decay  cone at each energy.
In practice, of course, $L$ and $R_d$ (which fix $\theta_d$) and $E_\mu$
depend on and are constrained by various factors like geographical location 
of existing facilities, physics goals, cost and design considerations. 
It is thus useful to examine the behaviour of the $\tau$ production
event rate when, for instance, $L$ and $R_d$ are fixed and $E_\mu$ is varied,
with a view to optimization.
 
\par
\begin{figure}[htb]
\vspace*{-1.9 in}
\centerline{\epsfig{file=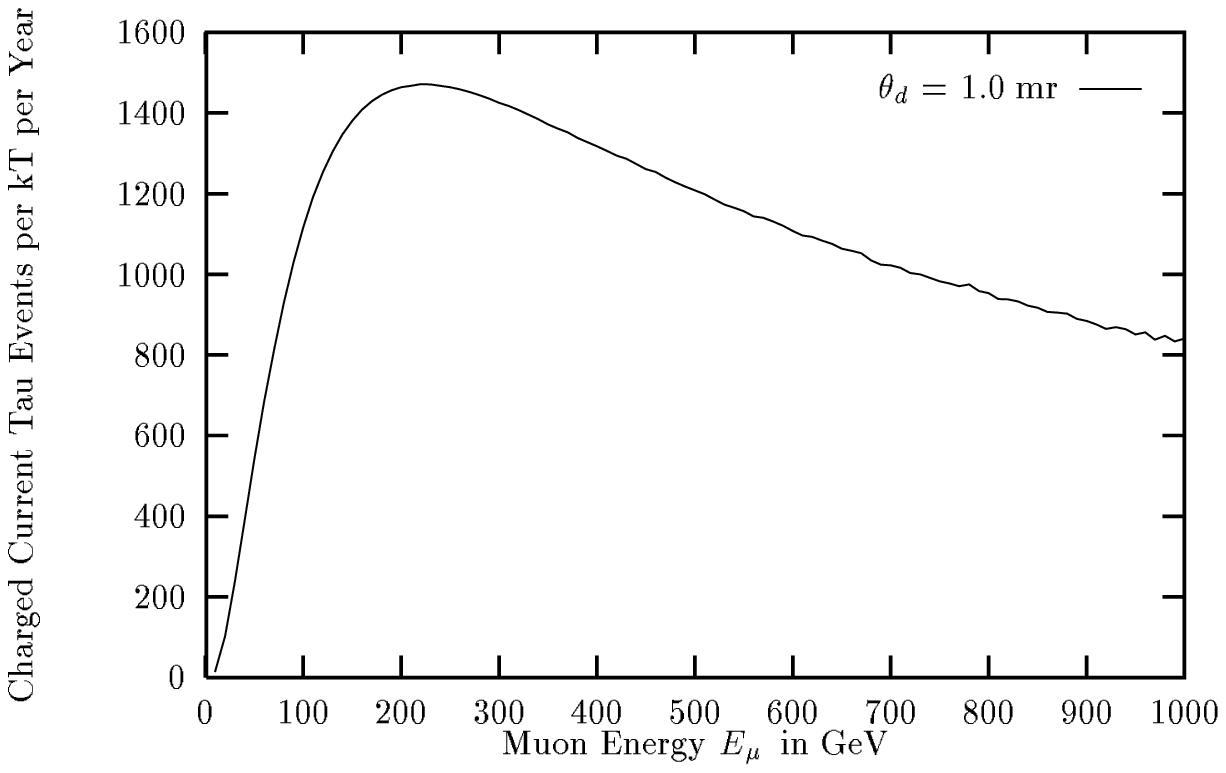,width=20cm}}
\vspace*{-6.6 in}
\caption{\em Variation of tau events with  muon beam energy for a
   baseline length of 732 km and $\Delta m^2 =.0022$ and
  $\sin^2\,2\theta = 1$} \label{fig4} 
\end{figure}

   
 In figure (\ref{fig4}) we plot the tau event-rate for $\theta_d = 10^{-3}$ radians, 
$\Delta m^2 =2.2 \times 10^{-3}$
eV$^2$, $\sin^22\theta = 1$ and baseline length L= 732 km. (The choice of 
baseline   is appropriate to a proposed beam from either CERN 
to Gran Sasso \cite{ngs}
or from Fermilab to the Soudan laboratory \cite{numi}.) 
After the initial rise with energy, it peaks around $E_\mu= 200$ GeV,
and then falls and flattens asymptotically with increasing energy. 
As the position of this peak with respect to $E_{\mu}$ depends on 
$\theta_d$ alone, it
 will remain invariant under changes of $\delta m^2$ and $\sin^22\theta$, as
long as Eq. (\ref{eq5}) remains a good approximation.

\par
\begin{figure}[htb]
\vspace*{-1.9 in}
\centerline{\epsfig{file=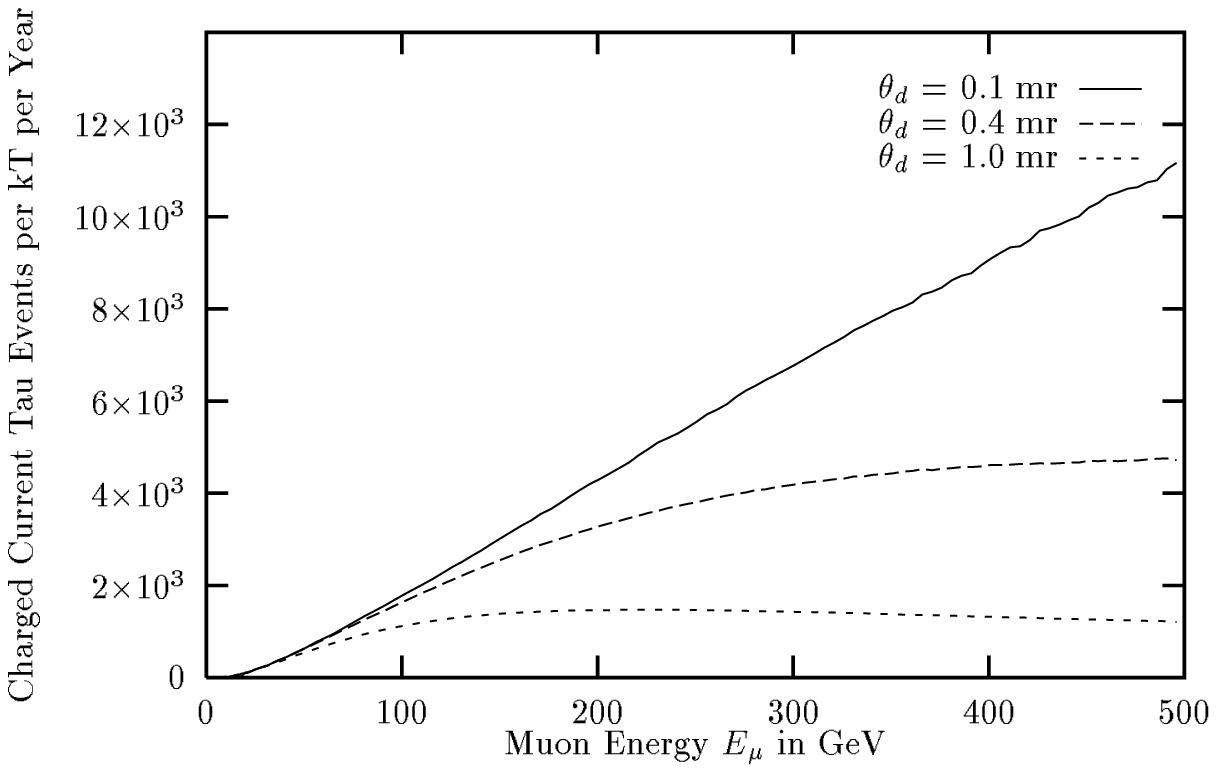,width=20cm}}
\vspace*{-6.6 in}
\caption{\em Variation of tau events for different detection cones 
with  muon beam energy for a
  given baseline length of 732 km and $\Delta m^2 =.0022$ and
  $\sin^2\,2\theta = 1$} \label{fig5} 
\end{figure}

\begin{figure}[htb]
\vspace*{-1.9 in}
\centerline{\epsfig{file=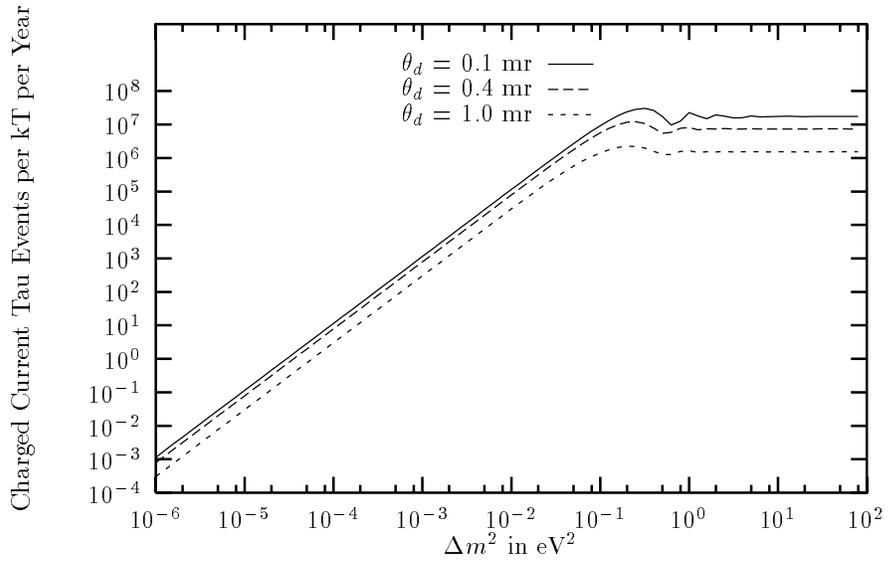,width=20cm}}
\vspace*{-6.6 in}
\caption{\em Variation of tau events for different detection cones 
with $\Delta m^2$ for  muon beam  energy of 250 GeV, 
   baseline length of 732 km and $\Delta m^2 =.0022$ and
  $\sin^2\,2\theta = 1$} \label{fig6} 
\end{figure}

   Figures \ref{fig5} and \ref{fig6} compare the yields per kT-yr for 
various detection cones, respectively, as beam energy and $\Delta m^2$ are
varied.     For low beam
energies where forward peaking of
the decay products is not pronounced, narrower detection cones contain
fewer events/kT/yr , but for higher 
energies the behaviour is reversed (figure \ref{fig5}). For instance, a
 detector which subtends a $\theta_d = 10^{-4}$ radians  cone
will see  roughly 11  times more events/kT/year for $E_\mu=500$ GeV than one subtending
$\theta_d = 10^{-3}$ radians, since it intercepts a higher
average flux $\left\langle\Phi_{\nu_\mu}
\right\rangle$. The assumption of a uniform flux over the area of the detector
can thus be misleading, since the event rate scales very non-linearly   with
the detector area.
In Figure (\ref{fig6}), the rise in the event-rate as $\Delta m^4$ signalled
by Equations (\ref{eq5}) and (\ref{eq1}) in conjunction is clearly apparent
upto $\Delta m^2$ values of $O(0.5)$ eV$^2$, after which the sinusoidal 
behaviour sets in. 

\par In conclusion, as illustrated by figures \ref{fig2} - \ref{fig6}, $\tau$ event rates from
$\nu_\mu$ beams at muon colliders have several interesting
characteristics which are relevant to experimental design and choice
of baseline length and beam energy:
\begin{enumerate}
\item For a substantial and physically interesting range of $\Delta
m^2$, the event rate (events/kT/year) for a fixed $\theta_d$ is
 to a very
good approximation independent of baseline length for a wide range of
beam energies (figure \ref{fig3} ).
\item For a given choice of say,  baseline and detector area
$({\it{i.e.}}$ fixed $\theta_d)$,  the  event rate
is maximised at a particular beam energy, independent of  the particular 
values  of $\Delta m^2$ and
$\sin^22\theta$ over a considerable portion of their range (figure \ref{fig4}).
\item The intense forward peaking expectedly renders detectors 
with smaller area
superior to those with large area at high beam energies, for a fixed 
available mass of the detector, and we show the extent to which 
this affects actual event rates in figure (\ref{fig5}). 
\end{enumerate}

\section{Selection of Tau and Wrong - Sign Muon events} 
An important component of any study for $\tau$ appearance due to 
$\nu_{\mu,e}  \rightarrow \nu_{\tau}$
  oscillations is the event selection
strategy for the $\tau$'s  produced  from charged current interactions
of the $\nu_{\tau}$.  This has been discussed in the literature in the
context  of several terrestrial experiments that are already 
in progress \cite{chorus}, \cite{nomad}. Strategies for
$\tau$-detection   have also received consideration in proposals  for future experiments \cite{icarus}, \cite{opera}.
\par For  neutrino experiments using a muon storage ring,
the detailed prescription for event selection can be formulated only
after the detector design is specified.  There are, however,
some basic issues concerning the signals and the backgrounds which
all  experiments are likely to be concerned with. We base our
predictions  here on these considerations, implemented within a
parton-level Monte Carlo calculation.  
\par  The results presented by us are in connection with a
  detector   of mass 10 kT similar to that described in \cite{geer},  placed
perpendicularly to the  muon beam axis, with a short, medium or large
baseline, incorporating detailed tracking and particle identification
facilities.  

\par We have based our first set of calculations on tau-detection through the
one-prong  hadronic decay channels. This essentially includes the decays
$\tau^{\pm}  \longrightarrow \pi ^{\pm} \nu_\tau$, 
$\tau^{\pm}  \longrightarrow \rho^{\pm} \nu_\tau$ and  
$\tau^{\pm}  \longrightarrow {a_1} ^{\pm} \nu_\tau$, with  small
additional contributions from the $K$ and $K^*$ channels
\cite{bullock}.  The $\rho^{\pm}$ subsequently decays into 
$\pi^{\pm} \pi^0$. 
For the decay of $a_1$, we have confined ourselves to the mode $\pi^{\pm}
{\pi^0} {\pi^0}$   which leads to a single charged-track.
The branching ratios into these channels \cite{pdg} are approximately
11$\%$, 25$\%$ and  9$\%$ respectively, giving a 
substantial total branching ratio
of about  $45\%$. Thus, essentially  one should look
for one charged pionic track with 0, 1 or 2 neutral pions in a
collinear  configuration. The  total energy, measured from deposits in
the  electro-magnetic and hadronic calorimeters  gives the energy of the
$\pi^{\pm}$, $\rho^{\pm}$ or ${a_1}^{\pm}$, which can be combined with
the  directional information to construct its three-momentum.
\par The backgrounds for signals of this kind (kinks in the charged
tracks  with missing $p_T$) can come, for example, from re-interaction
of the  hadronic jets coming out of the deep inelastic scattering (DIS) 
vertex, particularly in the case of neutral current events. Charmed
particle  production and decays in the DIS   processes  can also
give  rise to potential backgrounds.  There is also the possibility 
of muons (from  charged current events with no oscillation) being
misidentified  as pions. And finally, one can have the so-called
`white kinks'  arising from one-prong nuclear interactions with no
heavy ionising tracks. (These types of kinks usually have
small $p_T$,  within about 500 MeV for our energy
ranges.) 
\par With the above considerations in mind,  our
first set of  results (for a  10 kT detector) implements  the following
event  selection criteria for a 250 GeV muon beam \cite{dp}:
\begin{itemize}
\item A minimum $\not{p_T}$ of 0.5 GeV. 
\item a minimum energy of 2 GeV for the one-prong decay products from
the  tau's.
 \item A minimum  isolation of $\Delta R$ = 0.7 between the charged
prong from  tau-decay and the DIS products, where $\Delta R^2 = \Delta
\eta^2  + \Delta \phi^2$,  $\Delta \eta$ and  $\Delta \phi$ being
the differences in  pseudo-rapidity and azimuthal angle respectively. 
\end{itemize}
\par The last criterion  ensures that the one-prong charged tracks
characteristic of $\tau$-decays are at such angles with the beam axis
as to set them clearly apart from misidentified muons produced from
unoscillated $\nu_\mu$'s as well as from white kinks. At 
lower (higher) energies, the $\not{p_T}$ cut has to be slightly reduced 
(enhanced) in order to suppress backgrounds with the same effectiveness.

 \par We find that the missing-$p_T$ and isolation cuts
taken together  can remove the entire set
of backgrounds due to unoscillated charged-current events,  whereas
the  neutral current backgrounds are adequately taken care of by the
isolation  cuts. 
Taking  everything together, the approximate efficiency of tau
detection in our  parton-level calculation turns out to be
 31$\%$ (including the branching ratio for one-prong decays). 
This is commensurate with the efficiencies expected in, for example,
the  OPERA experiment \cite{opera}. 
\par Our second set of results is based on tau appearance due to $\overline{\nu_e}$
oscillating  into  $\overline{\nu_\tau}$.
 These
will lead to wrong-sign muons via charged current interactions. Such
signals  are relatively background-free;  the only significant
backgrounds   come  from charm production at the DIS vertex. 
 For this set of our results we have
\begin{itemize}
\item A minimum $\not{p_T}$ of 0.2 GeV.
\item a minimum energy of 2 GeV for the wrong sign charged lepton decaying from tau
(muon and electron).
\item A minimum transverse momentum of wrong sign muon/electron $P_T^{\mu , \, e}$ 
of 0.2 GeV
 \item A minimum  isolation of $\Delta R$ = 0.4 between the wrong sign charged
lepton from  tau-decay and the DIS products.
\end{itemize}

\par Finally, although our focus here is on $\tau$ appearance, we also 
give results in Section 4 for the detection of $\overline{\nu_e}
\rightarrow  \overline{\nu_\mu}$ oscilations, detectable again by 
the presence of wrong sign muons, after implementing suitable cuts.
 
\section{ Contours  for $\nu_{\mu,e}\rightarrow \nu_{\tau}$
 oscillation searches at muon storage rings}

\par In order to demonstrate the possibilities offered by muon storage rings
for oscillation studies, we give the corresponding $90\%$ C.L. contours for 
$ \Delta m^2 $ and $ \sin^22\theta $ for two-flavour mixing.
As mentioned earlier, we feel this is adequate at
 present to obtain a firm feel for the 
eventual potential of these experiments to comprehensively map the parameter 
space of neutrino masses and mixing.

\par Starting with  equation (\ref{eq1}) for $N_\tau$, the \lq bare \rq
events, and folding in the kinematic cuts described above for event selection
and background elimination, one obtains  $N^d_\tau$, 
representing the actual candidate events. Requiring $N^d_\tau \leq 2.44$ 
then delineates the $90\%$ C.L. parameter space. Thus, for each contour 
the average value of the probability
\begin{equation}
 {\bar{\cal P}}_{\nu_{\mu,e}\rightarrow \nu_{\tau}} = 2.44/
N_\tau^{all} \label{eq6}
\end{equation}
where $N_\tau^{all}$
 is computed from equation ( \ref{eq1}) by setting 
$ {\cal P}_{\nu_{\mu,e}\rightarrow \nu_{\tau}} = 1$, representing
total conversion of $\nu_\mu$'s to $\nu_\tau$'s, 
but imposing the cuts as before.
Also, we note that \cite{deruj} for each contour, the reach in $\Delta m^2$, 
{\it{i.e.}} its minimum value, occurring when $\sin^22\theta = 1$ is 
given to a good approximation by 
\begin{equation}
 {\Delta m^2}_{min} = {\left(\frac{2.44}{ N_\tau^{all}}\right)}^{1/2} 
\frac{\langle E_\nu \rangle}{L}.\label{eq7}
\end{equation}
Since $N_\tau^{all}$ scales as the product of the flux and the cross-section,
(having the probability term set to 1), it follows that
\begin{equation}
 {\Delta m^2}_{min} \propto \langle E_\nu\rangle^{-1/2} .\label{eq8}
\end{equation}
 Similarly, the \lq knee \rq for each contour,
{\it{i.e.}} the minimum value of $\sin^22\theta$ probed occurs when the 
other oscillating term in equation (\ref{eq2}) is approximately 1, hence
\begin{equation}
 {\sin^22\theta}_{min} \approx {\bar{\cal P}}_{\nu_{\mu,e}\rightarrow 
\nu_{\tau}}.\label{eq9}
\end{equation}
which implies that 
\begin{equation}
 {\sin^22\theta}_{min} \propto \frac{ L^2}{ 
\langle E_\nu\rangle^{3}}.\label{eq10}
\end{equation}
 Finally, the vertical asymptotic part of the contour, occurring when the 
values of $\Delta m^2 $ are high enough such that the sine squared 
term containing it in equation (\ref{eq2}) averages to $1/2$ has
\begin{equation} 
\sin^22\theta = 2{\bar {\cal P}}_{\nu_{\mu,e}\rightarrow \nu_{\tau}}.
\end{equation}

\par
\begin{figure}[htb]
\vspace*{-1.9 in}
\centerline{\epsfig{file=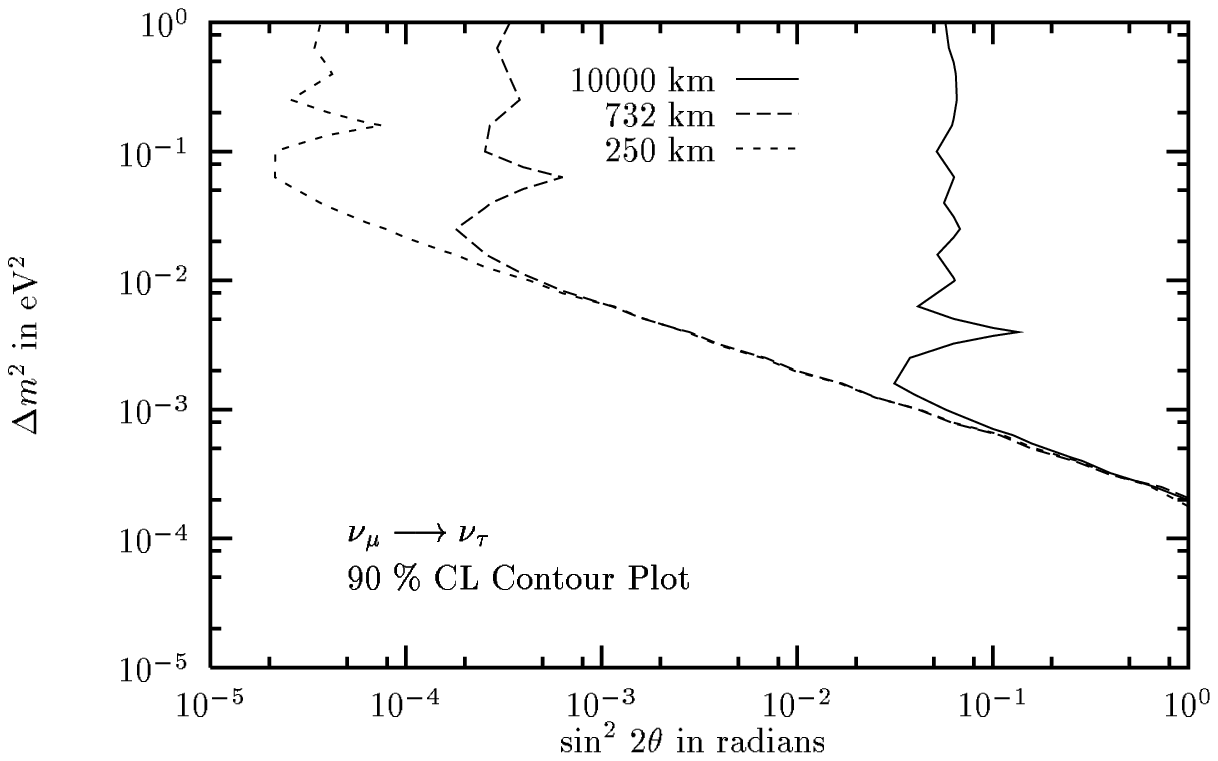,width=20cm}}
\vspace*{-6.6 in}
\caption{\em $90\%$ CL $\nu_\mu \longrightarrow \nu_\tau$ 
oscillation contour plot  for 20 GeV,  $2\times 10^{20}$ Muons 
per year with 10 kT
target and  angular opening of  0.1 milli radian. $\tau$ detection is
through one prong hadronic decay (see text)  } 
\label{fig7}
\end{figure}

\begin{figure}[hbt]
\vspace*{-1.9 in}
\centerline{\epsfig{file=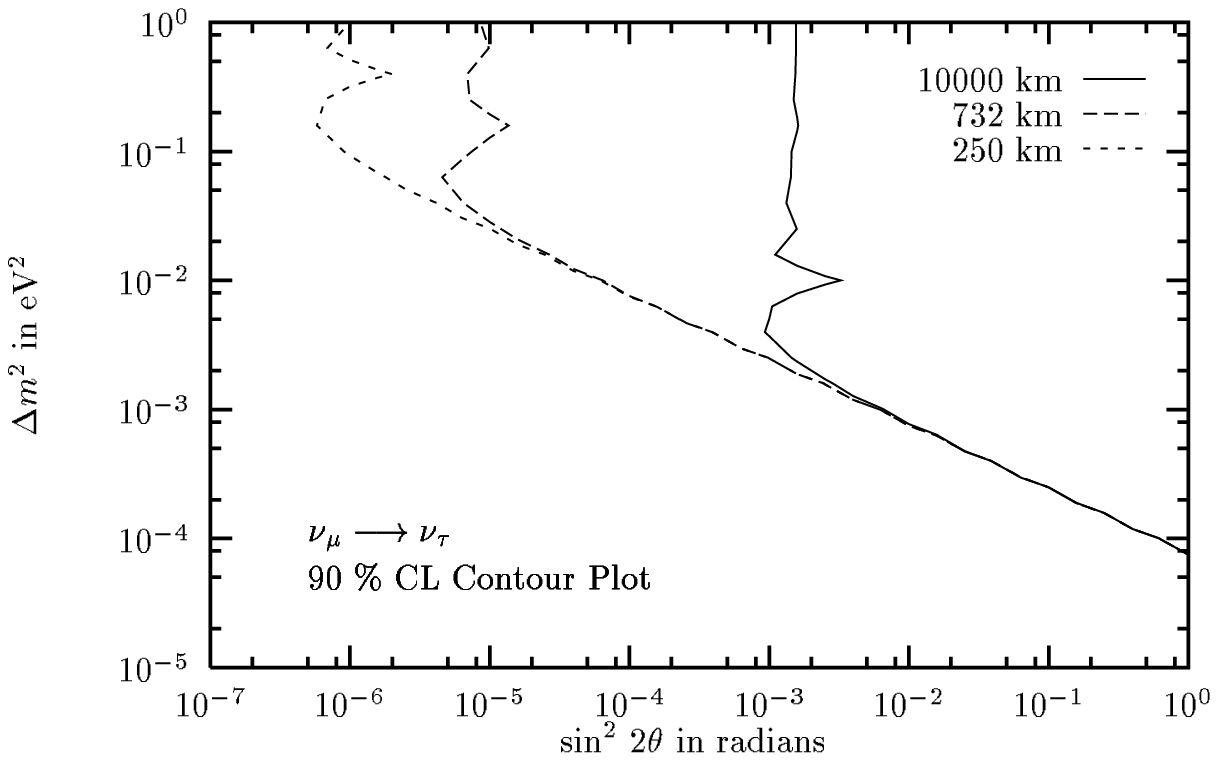,width=20cm}}
\vspace*{-6.6 in}
\caption{ \em $90\%$ CL $\nu_\mu \longrightarrow \nu_\tau$ 
oscillation  contour plot  for 50 GeV,  $2\times 10^{20}$ Muons 
per year with 10 kT
target and  angular opening of  0.1 milli radian.  $\tau$ detection is
through one prong hadronic decay.
 (see text) } 
\label{fig8}
\end{figure}

\begin{figure}[htb]
\vspace*{-1.9 in}
\centerline{\epsfig{file=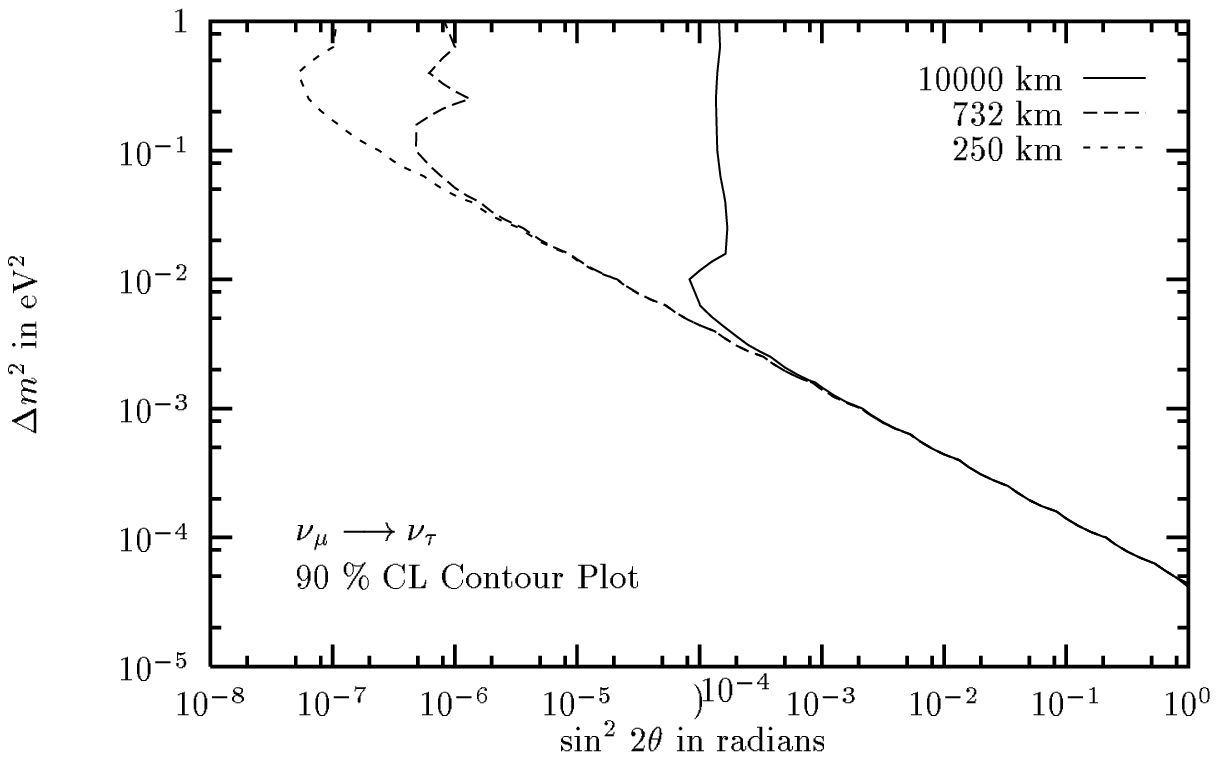,width=20cm}}
\vspace*{-6.6 in}
\caption{\em  $90\%$ CL $\nu_\mu \longrightarrow \nu_\tau$ 
oscillation contour plot  for 100 GeV,  $2\times 10^{20}$ Muons 
per year with 10 kT
target and  angular opening of  0.1 milli radian. $\tau$ detection is
through one prong hadronic decay (see text)  } \label{fig9}
\end{figure}

\begin{figure}[htb]
\vspace*{-1.9 in}
\centerline{\epsfig{file=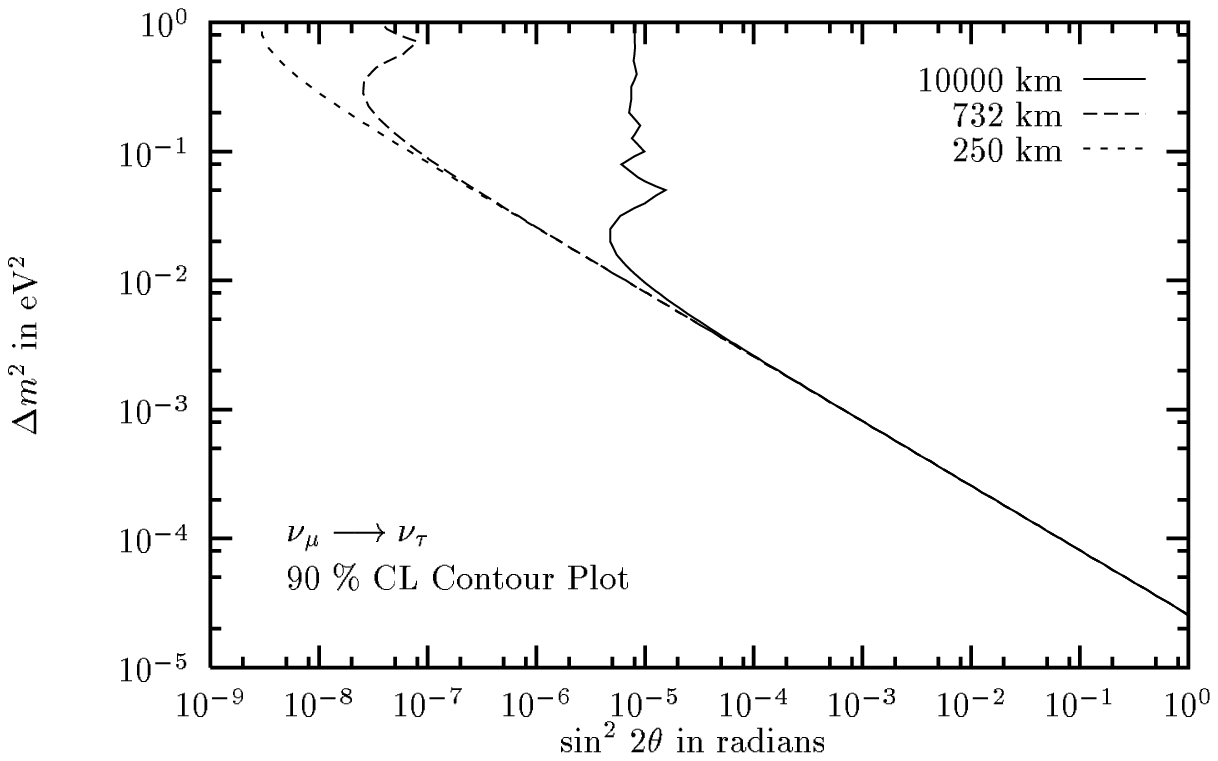,width=20cm}}
\vspace*{-6.6 in}
\caption{ \em $90\%$ CL $\nu_\mu \longrightarrow \nu_\tau$ 
oscillation  contour plot  for 250 GeV,  $2\times 10^{20}$ Muons 
per year with 10 kT
target and  angular opening of  0.1 milli radian. $\tau$ detection is
through one prong hadronic decay (see text).  } \label{fig10}
\end{figure}

\begin{figure}[htb]
\vspace*{-1.9 in}
\centerline{\epsfig{file=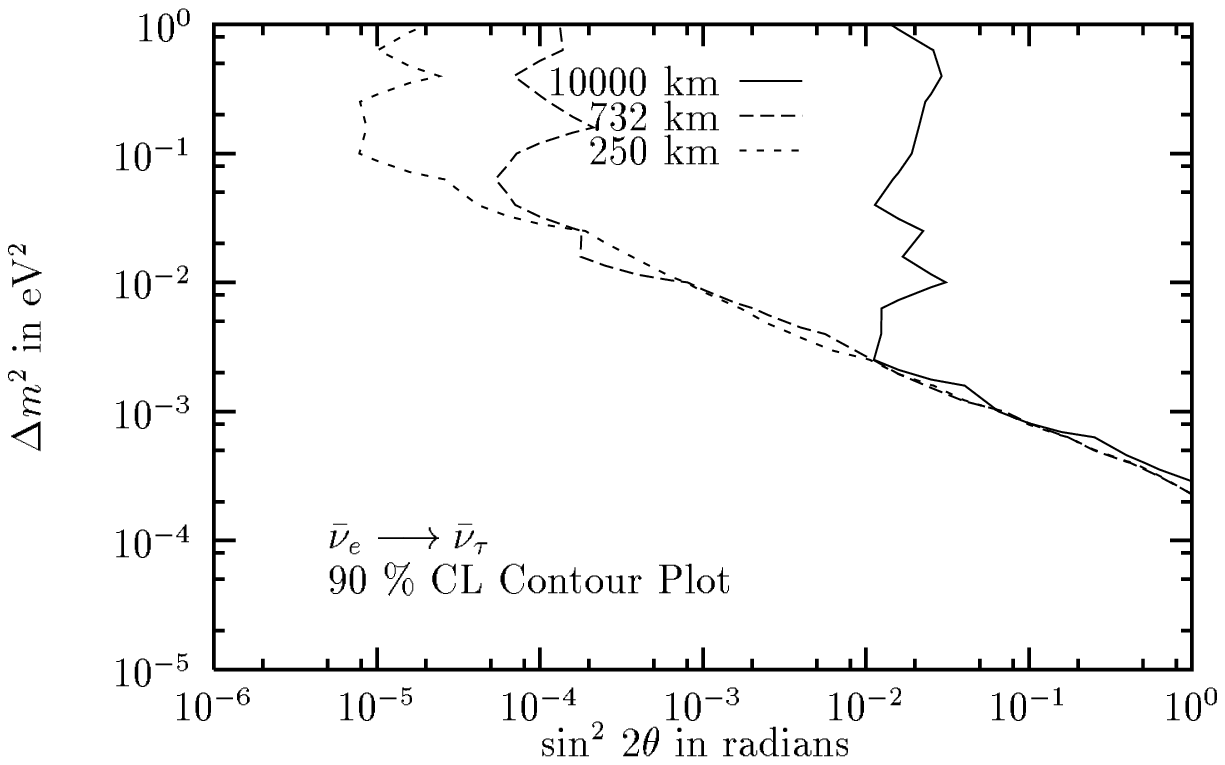,width=20cm}}
\vspace*{-6.6 in}
\caption{\em $90\%$ CL $\bar\nu_e \longrightarrow \bar\nu_\tau$ 
oscillation contour plot  for 50 GeV,  $2\times 10^{20}$ Muons 
per year with 10 kT
target and  angular opening of  0.1 milli radian. $\tau$  detection 
through the  leptonic decay mode (see text)  } \label{fig11}
\end{figure}

\begin{figure}[htb]
\vspace*{-1.9 in}
\centerline{\epsfig{file=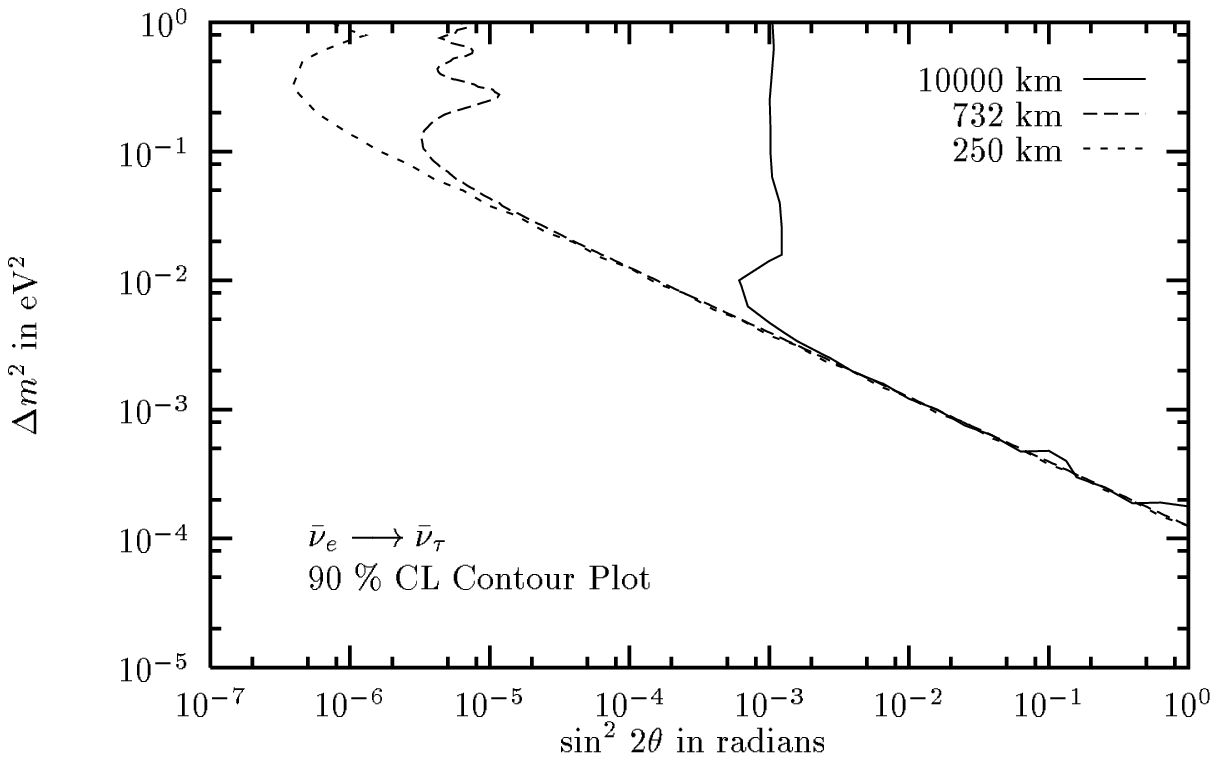,width=20cm}}
\vspace*{-6.6 in}
\caption{ \em $90\%$  CL  $\bar\nu_e \longrightarrow \bar\nu_\tau$ 
oscillation  contour plot  for 100 GeV,  $2\times 10^{20}$ Muons 
per year with 10 kT
target and  angular opening of  0.1 milli radian. $\tau$ detection 
through  the leptonic decay mode (see text)  } \label{fig12}
\end{figure}

\begin{figure}[htb]
\vspace*{-1.9 in}
\centerline{\epsfig{file=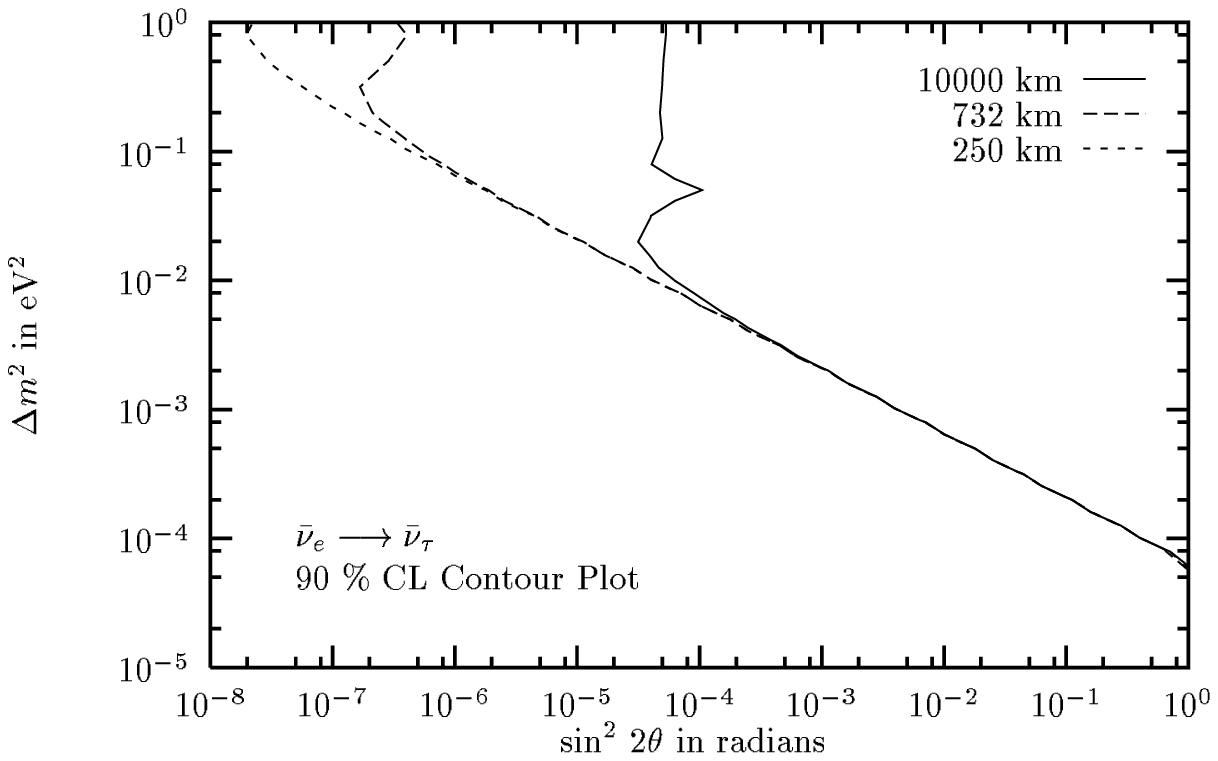,width=20cm}}
\vspace*{-6.6 in}
\caption{ \em $90\%$ CL  $\bar\nu_e \longrightarrow \bar\nu_\tau$ oscillation  contour plot  for 250 GeV,  $2\times 10^{20}$ Muons 
per year with 10 kT
target and  angular opening of  0.1 milli radian. $\tau$  detection 
through  the leptonic decay mode (see text)  } \label{fig13}
\end{figure}

 Figures \ref{fig7},\ref{fig8},\ref{fig9} and \ref{fig10} show contours 
for $\tau$ lepton detection via 
decay modes  containing hadronic final states, for muon beam energies of
$20$ GeV, $50$ GeV, $100$ and $250$ GeV at $90\%$ C.L. 
for a 10 kT-yr detector,
assuming a detection cone of $0.1$ milli radians. 
Each figure has contours for 
three different baselines, 
$250$ km (Kamioka to KEK), $732$ km (CERN to Gran Sasso or
Fermilab to Soudan) and $10000$ km (Fermilab to Japan). Figures 
\ref{fig11},\ref{fig12} and \ref{fig13}
show curves for the same choices of baselines and other parameters, but for 
energies of $50$ GeV, $100$ GeV and 250 GeV and for the wrong sign lepton detection
for $\bar{\nu_e} \rightarrow \bar{\nu_\tau}$ oscillation. For a muon beam 
energy of $20$ GeV, no wrong-sign muon events 
survive our (conservative) cuts. 

\par
In the regions corresponding to  small mass-squared differences, all the contours
for a given energy tend to merge. This again demonstrates the insensitivity 
to baseline in these regions. The scaling relation for
$\sin^2\theta_{min}$ with length (for a fixed beam energy) ( equation (\ref{eq10}) ) 
is also reproduced well in the contours. Once one compares different energies,
however, the scaling is distorted by the presence of energy-dependant cuts
and also marginally  by the fact that the plots here are for muon beam energies, while 
the scaling relations are for neutrino energies. It is apparent from Equations 
(\ref{eq8}) and (\ref{eq10}) and from the contours that very long baselines
offer no advantages for oscillation studies employing neutrinos from muon
storage rings, and carry the added burden of impractical detector sizes,
at least if conventional detectors are employed \footnote{Our reason for 
focussing on medium and long baselines is in part due to the geography 
of existing sites where such experiments may be possible, and in part due to 
the fact that short baselines may require new and different detectors, which
are currently under active study, as discussed in \cite{king,deb}}. 
 
The dramatic potential for oscillation studies is apparent.  Even a modest
$20$ GeV beam energy is capable of effectively mapping the 
$\nu_\mu \rightarrow \nu_\tau$ parameter space indicated by Super K, 
{\it{i.e.}} $\Delta m^2 \sim 10^{-3}- 5 \times 10^{-2}$ and 
$\sin^22\theta \sim 0.8 - 1$, as can be seen  from Fig.\ref{fig7}.
 As expected from equation (\ref{eq10}), shorter baselines provide an enhanced 
reach in $\sin^22\theta_{min}$
 ($\propto L^2$), for the same beam energy and reach in $\Delta m^2$.   
For  $\nu_\mu \rightarrow \nu_\tau$ searches, muon storage rings provide
the potential to greatly exceed limits set by present and past 
accelerator experiments (NOMAD, CHORUS, CDHS, CCFR and E531) \cite{con}. 
At present, bounds on
this oscillation mode cover the range $\sin^22\theta \sim 10^{-3} - 1$  and 
$\Delta m^2 \geq 1$ eV$^2$. Beam energies of $100$ GeV and a baseline
of 250 km would allow exploration upto $\sin^22\theta_{min} \sim 10^{-7}$
and $\Delta m^2_{min} \sim 5 \times 10^{-5}$ (Figure \ref{fig9}).
 Similarly, the 
wrong-sign muon mode contours (figures \ref{fig11},\ref{fig12} and \ref{fig13} )
 demonstrate the potential for 
extending the limits explored for 
$\nu_e \rightarrow \nu_\tau$ parameters. Current bounds from 
accelerator and reactor based experiments (NOMAD, CCFR, BUGEY
and CHOOZ) \cite{con} go down to $\Delta m^2 \geq 10^{-3}$ eV$^2$ and
$\sin^22\theta \sim 5 \times 10^{-2} - 1$. A 100 GeV muon beam and a 250 km
baseline would extend that to $2 \times 10^{-4}$ eV$^2$ and $5 \times 10^{-7}$
respectively.

\begin{figure}[htb]
\vspace*{-1.9 in}
\centerline{\epsfig{file=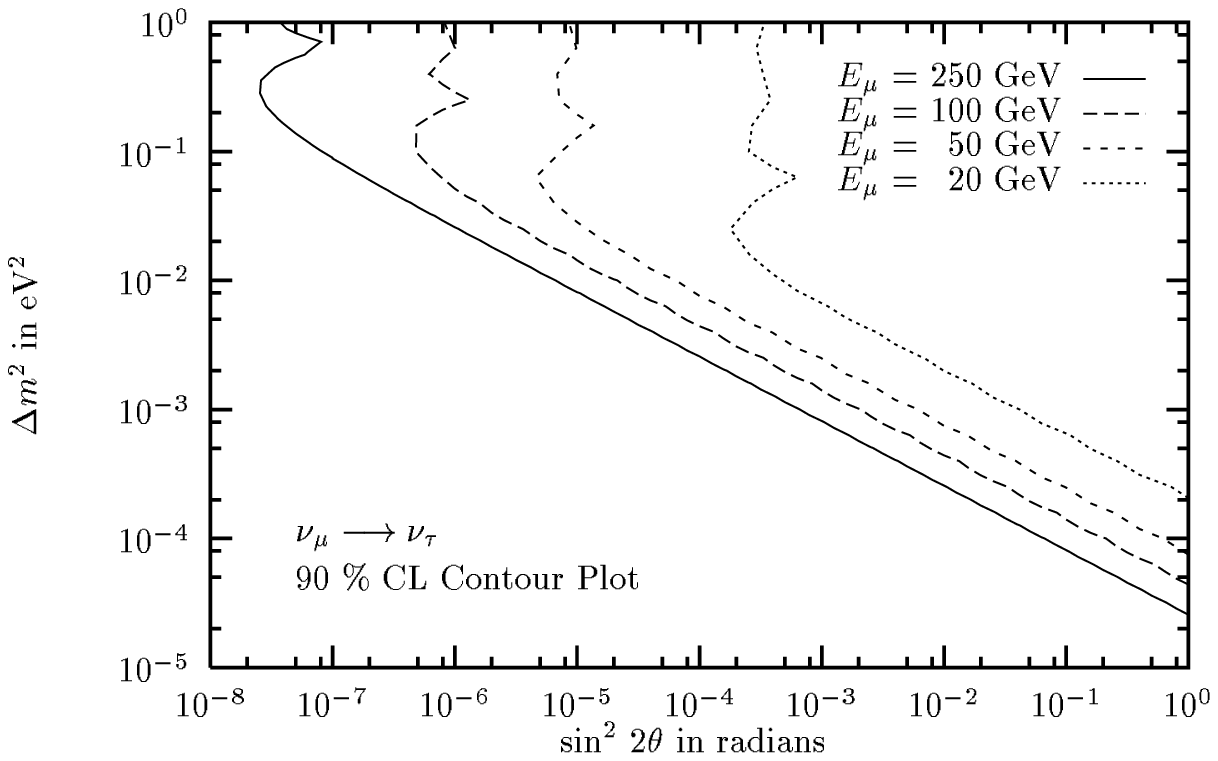,width=20cm}}
\vspace*{-6.6 in}
\caption{\em $90\%$ CL $\nu_\mu \longrightarrow \nu_\tau$ 
oscillation  contour plots for 732 km. baseline experiment,
and  $2\times 10^{20}$ Muons per year with 10 kT
target and  angular opening of  0.1 milli radian. $\tau$ detection 
through  one prong hadronic decay mode (see text). } \label{fig14}
\end{figure}

\begin{figure}[htb]
\vspace*{-1.9 in}
\centerline{\epsfig{file=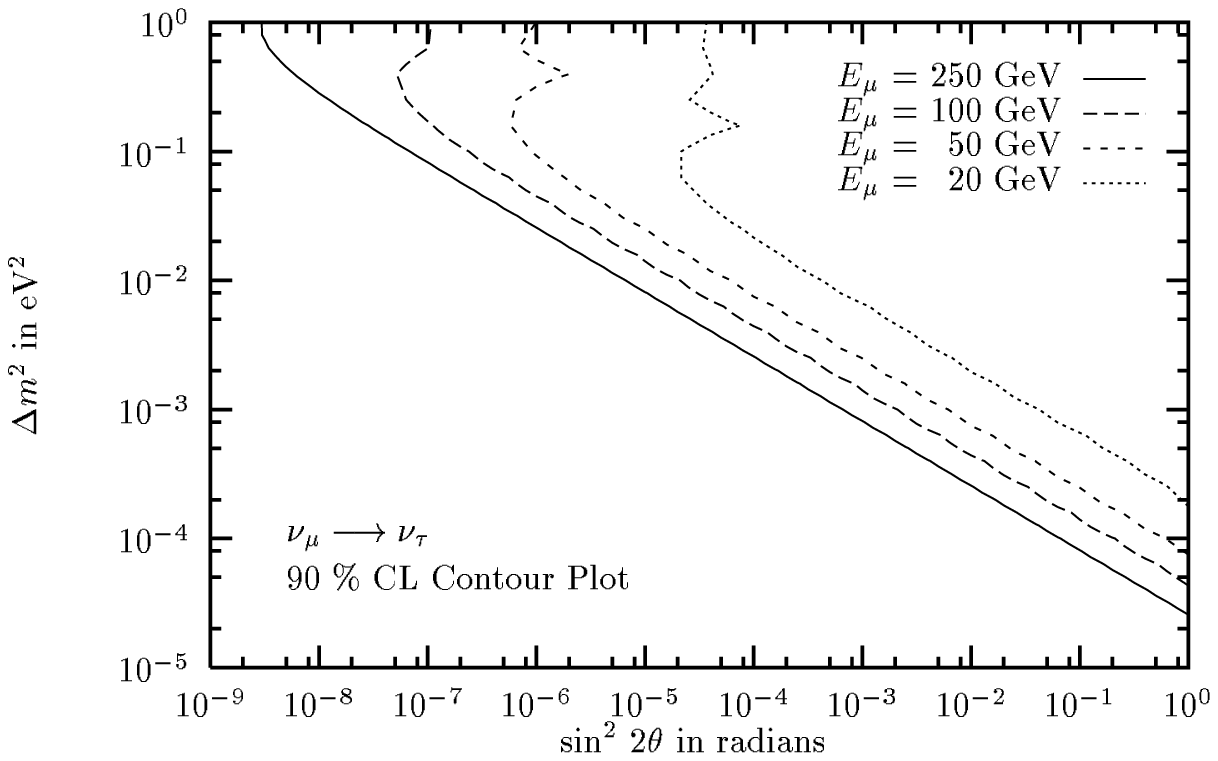,width=20cm}}
\vspace*{-6.6 in}
\caption{ \em $90\%$ CL $\nu_\mu \longrightarrow \nu_\tau$ 
oscillation  contour plots for 250 km. baseline experiment,
and  $2\times 10^{20}$ Muons per year with 10 kT
target and  angular opening of  0.1 milli radian. $\tau$ detection 
through  the one prong hadronic decay mode (see text).  } \label{fig15}
\end{figure}

\begin{figure}[htb]
\vspace*{-1.9 in}
\centerline{\epsfig{file=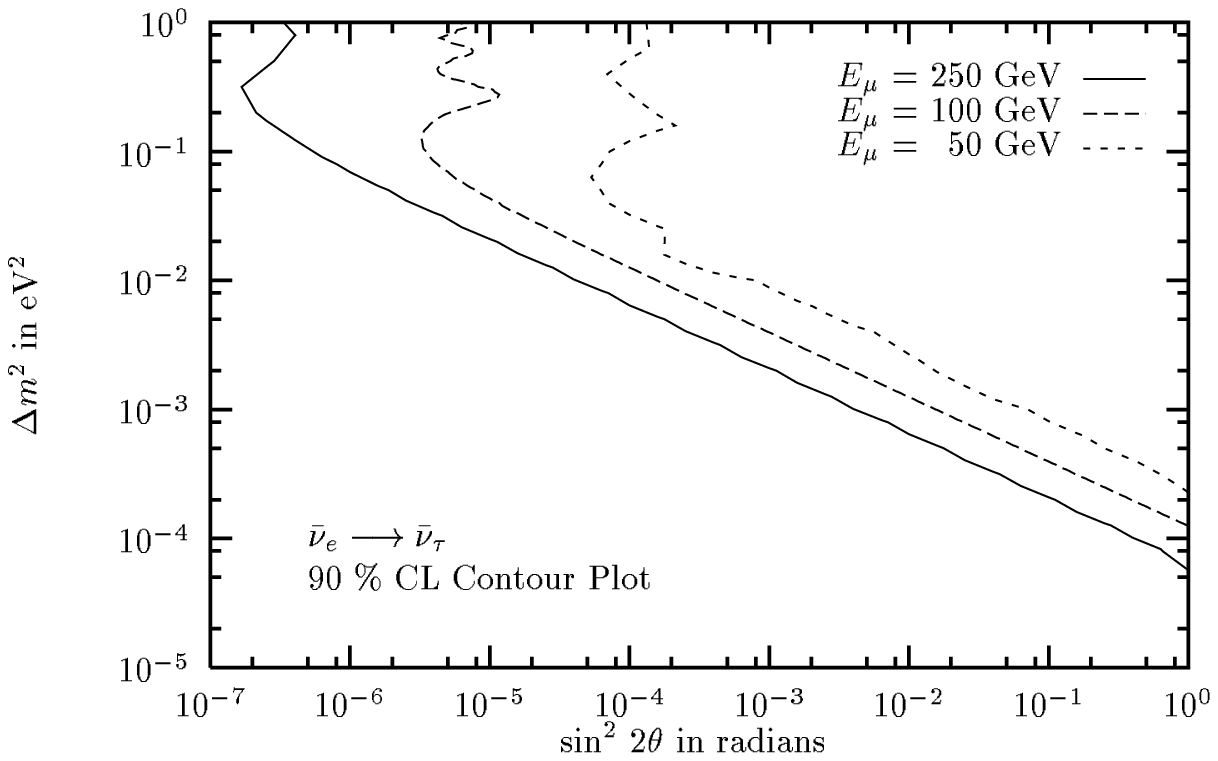,width=20cm}}
\vspace*{-6.6 in}
\caption{\em  $90\%$ CL $\bar\nu_e \longrightarrow \bar\nu_\tau$ 
oscillation  contour plots for 732 km. baseline experiment,
and  $2\times 10^{20}$ Muons per year with 10 kT
target and  angular opening of  0.1 milli radian. $\tau$ detection 
through the  leptonic decay mode (see text). }  
   \label{fig16}
\end{figure}

\begin{figure}[htb]
\vspace*{-1.9 in}
\centerline{\epsfig{file=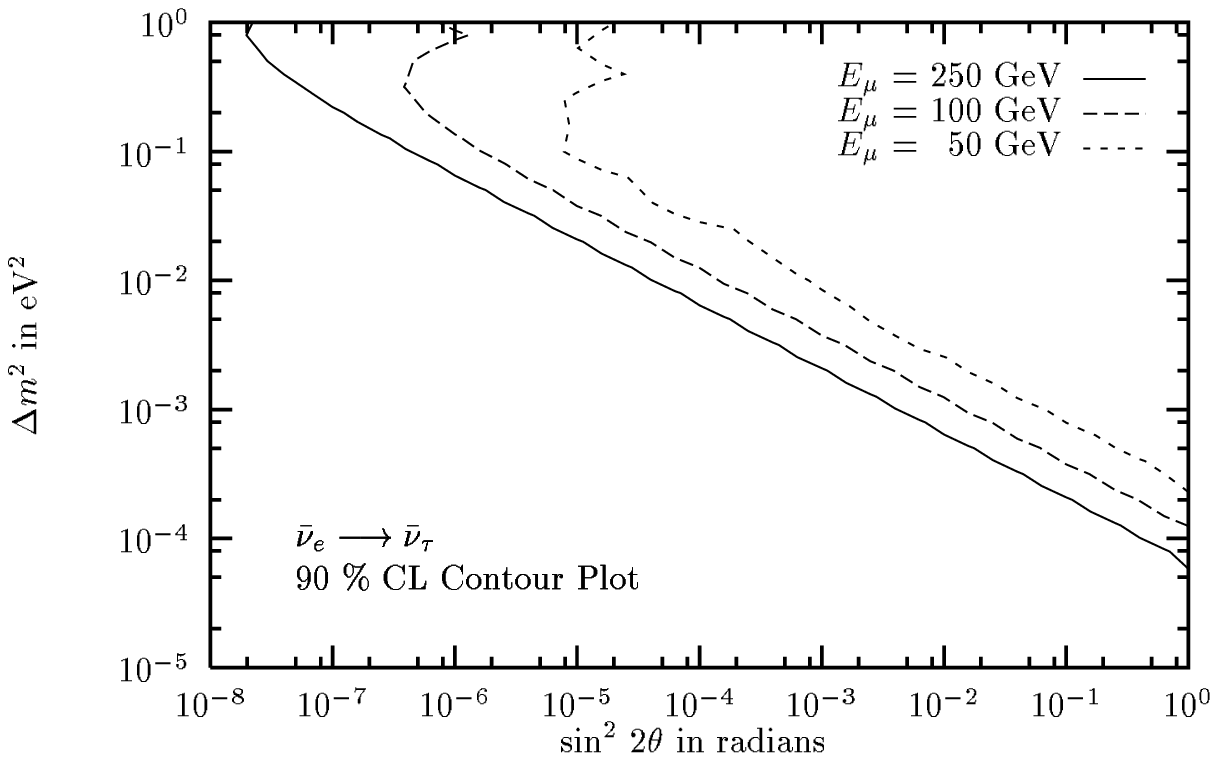,width=20cm}}
\vspace*{-6.6 in}
\caption{\em  $90\%$ CL $\bar\nu_e \longrightarrow \bar\nu_\tau$ 
oscillation  contour plots for 250 km. baseline experiment,
and  $2\times 10^{20}$ Muons per year with 10 kT
target and  angular opening of  0.1 milli radian. $\tau$'s are detected 
through  leptonic decay mode (see text). }  
\label{fig17}
\end{figure}

For convenience, in figures \ref{fig14} -\ref{fig17}, we reproduce our contours showing how 
different baselines (250 km and 732 km) compare for various energies for both modes 
of tau appearance discussed above. The distortion of the 
scaling relations equations  (\ref{eq8}) and (\ref{eq10}) due to the presence
of experimental cuts is apparent here, being more pronounced at low 
beam energies. For 20 GeV and 50 GeV, (Figure (\ref{fig14})) for instance, $\Delta m^2_{min}$
scales as $\sim 1/E_\nu$ rather than $1/{\sqrt E_\nu}$.

\begin{figure}[htb]
\vspace*{-1.9 in}
\centerline{\epsfig{file=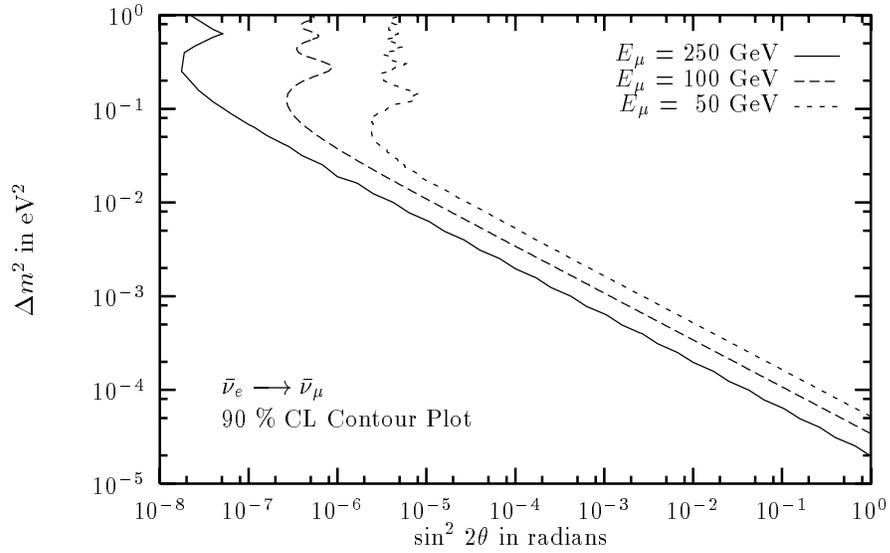,width=20cm}}
\vspace*{-6.6 in}
\caption{\em  $90\%$ CL $\bar\nu_e \longrightarrow \bar\nu_\mu$ 
oscillation  contour plots for 732 km. baseline experiment,
and  $2\times 10^{20}$ Muons per year with 10 kT
target and  angular opening of  0.1 milli radian.}  
   \label{fig18}
\end{figure}

Finally, although our focus here has been the detection of $\nu_\tau$
appearance, the experiments discussed here are in a natural position to 
also study $\overline{\nu_e}
\rightarrow  \overline{\nu_\mu}$ oscillations. The parameter regions
which can be explored are shown in figure \ref{fig18}, after incorporating
appropriate cuts to remove backgrounds for the wrong-sign muons. Clearly, the region identified
by the LSND experiment \cite{lsnd} can be scrutinised with ease at muon
storage rings.

\section{Conclusions}

We have studied the possibilities offered by muon storage rings 
(at various muon beam energies and baselines ) for $\nu_\tau$ appearance
experiments in order to determine masses and mixing angles for $\nu_\mu
\rightarrow \nu_\tau$ and $\nu_e
\rightarrow \nu_\tau$ oscillations. Tau event rates for such
experiments have first been discussed with a view to understanding their 
variation prior to the inclusion of experimental cuts, in order to 
better understand how baselines, forward peaking of decay 
neutrinos with increasing energies,
and average fluxes intercepted by detectors of
 various sizes can affect their optimization. 
Subsequently, event rates implementing cuts for hadronic and 
wrong sign leptonic modes are computed and used to plot contours for 
the parameter regions that can be explored in such experiments, and the 
expected scaling of the contours with energy and baseline is discussed.
 The results show that even for modest muon beam energies,
 convincing coverage and verification of the 
Super K parameters is possible. In addition, very significant
enlargement of present day bounds on the mixing parameters for oscillations 
to $\nu_\tau$ is guaranteed by these types of searches. 

\par In summary, neutrinos from muon storage rings appear to be  ideal sources,
 providing
 unprecedented potential for 
oscillation studies in the next millennium and 
deserve very serious consideration.
\section{Acknowledgement}
We  acknowledge useful conversations with D.P. Roy, D. Choudhury and Probir Roy  
and  also thank  S. Geer and B. King for helpful exchanges over e-mail. RG would
like to thank the CERN Theory Division for hospitality while this work was in progress.

\vfill
\eject

\clearpage

\begin{center}
\begin{table}[htb]
\begin{tabular}{||c c c c||}\hline
&{\bf CC Interaction and}  & {\bf Nature of fit} & {\bf Expression} \\ 
& $E_\nu$ {\bf Range}  & & \\ \hline
& & & \\
1.&$\nu_\tau  +\,\, {\rm N} \longrightarrow \tau^- +\,\, X$  &
 cubic & $-9.8656\times 10^{-8}\,E_{\nu_\tau}^3   +\,\, 2.5523\times 10^{-5}\,
E_{\nu_\tau}^2\,$\\ 
& & & \\
&20 - 100  GeV& & $+\,\, 4.1203\times 10^{-3} \,E_{\nu_\tau}   -   4.4066\times 10^{-2} $  \\ 
& & & \\
& & &  \\
2.& $\nu_\tau  +\,\, {\rm N} \longrightarrow \tau^- +\,\, X$  &
  linear &  $6.6685 \times 10^{-3}E_{\nu_\tau}   - 1.1579692\times 10^{-1}  $\\ 
& & & \\
& 50 - 2000  GeV& & \\ 
& &  & \\
&& & \\
3.& $\nu_\mu  +\,\, {\rm N} \longrightarrow \mu^- +\,\, X$  
 & cubic & $ 3.5953\times 10^{-11}\,E_{\nu_\mu}^3 -1.6794\times 10^{-7}\,
E_{\nu_\mu}^2\, $ \\
&& & \\
& 15  - 2000 GeV & & $ +\,\, 7.0057 \times 10^{-3}\,E_{\nu_\mu}   \,  +\,\, 2.6235\times  10^{-5}$ \\ 
& &  & \\ \hline
\end{tabular}
\caption{ Analytical expressions for the  fits of charge current
$\nu_\tau$ ($\nu_\mu$) cross-section in picobarns. $E_{\nu_\tau}$ ($E_{\nu_\mu}$) 
is in units of GeV.}
\label{table1}
\end{table}
\end{center}

\clearpage

\end{document}